\documentclass[a4paper,11pt]{article}

\usepackage{graphicx}
\usepackage{multirow}
\usepackage{makecell}
\usepackage[linesnumbered,ruled,vlined]{algorithm2e}
\usepackage{url}
\usepackage{hyperref}
\usepackage{comment}
\usepackage{xcolor}
\usepackage[english]{babel} 
\usepackage{soul}
\usepackage{booktabs}

\usepackage[left=2.5cm,right=2.5cm,top=3cm,bottom=2.5cm]{geometry}

\newcommand{\ignore}[1]{}  

\newcommand{\provisional}[1]{\textcolor{cyan}{#1}}

\title{UKFin+: A Research Agenda for Financial Services}
\author{Jing Chen \and Karen Elliott \and William Knottenbelt \and Aad van Moorsel \and Helen Orpin \and Sheena Robertson \and John Vines \and Katinka Wolter}

\date{\today}

\begin{document}
\renewcommand{\familydefault}{\sfdefault}
\sf
\maketitle


\section{Executive Summary}

This document presents a research agenda for financial services as a deliverable of UKFin+\footnote{See \url{https://ukfin.network}}, a Network Plus grant funded by the Engineering and Physical Sciences Research Council. UKFin+ fosters research collaborations between academic and non-academic partners directed at tackling complex long-term challenges relevant to the UK's financial services sector. Confronting these challenges is crucial to promote the long-term health and international competitiveness of the UK's financial services industry. As one route to impact, UKFin+ includes dedicated funding streams for research collaborations between academic researchers and non-academic organisations.

The intended audience of this document includes researchers based in academia, academic funders, as well as practitioners based in industry, regulators, charities or NGOs. It is not intended to be comprehensive or exhaustive in scope but may provide applicants to UKFin+ funding streams and other funding bodies with inspiration for their proposals or at least an understanding of how their proposals align with the broader needs of the UK financial services industry.

We identify six primary areas of real-life financial services challenges:
\vspace*{-2mm}
\begin{enumerate}
\vspace*{-2mm}
\item {\bf Financial Inclusion \& Wellbeing}: ensuring everyone has access to formal financial services on fair terms, and enabling everyone to feel secure in terms of their finances.
\vspace*{-2mm}
\item {\bf Coping with Game-Changing Technologies}: leveraging waves of emerging technologies with disruptive potential (especially AI, Blockchain, Quantum Computing).
\vspace*{-2mm}
\item {\bf Regulatory Compliance}: treading the fine line between stakeholder protection and stifling innovation.
\vspace*{-2mm}
\item {\bf Trusted \& Effective Open Banking}: fulfilling the potential of Open Banking and Open Data initiatives by increasing transparency and boosting public trust.
\vspace*{-2mm}
\item {\bf Financial Resilience \& Stability}: monitoring and modelling risk in complex, open and evolving market systems.
\vspace*{-2mm}
\item {\bf Sustainable Finance}: rating and integrating Environmental, Social and Governance issues into financial decisions.
\end{enumerate}

\vspace*{-2mm}
Finally, underpinning enablers for effective collaborations in all of these areas are highlighted.

\section{About UKFin+}

UKFin+ is a \pounds2.5m Network Plus grant funded by the Engineering and Physical Sciences Research Council (EPSRC). Its aim is to bridge the gap between university research and the needs of the financial services industry, consumers, and regulators. This gap has arisen at least partly because of an historical tendency to under-leverage academic investment: while academics focus on research and knowledge generation through new theories and models, industry is overwhelmingly focused on the application of knowledge through user-facing products and services.  

The specific focus of UKFin+ on the application of theories and models to the so-called {\bf wicked problems} of financial services. Wicked problems are complex and interdependent and resist singular solutions~\cite{ritchey2013}. Our society is filled with many such problems related to financial services and digital technology, e.g.\ financial inclusion, sustainability, cybersecurity for vulnerable populations, governance and prediction of economic markets. To be effectively addressed, wicked problems like these need multiple responses and collaborations across sectors and disciplines.

The core of UKFin+ features a \pounds1m commissioning framework that provides funding for different kinds of collaborative research projects aligned with UKFin+ priorities ranging from speculative {\em agile} projects to more mature {\em feasibility} projects\footnote{See \url{https://ukfin.network/funding-streams/}.}.  The UKFin+ commissioning framework also provides funding to strengthen enablers e.g. collaboration workshops and skills development.

\section{National Context}

The health of the UK economy is critically dependent on its services sector, which is responsible for over 80\% of the UK's economic output and over 80\% of employment in Q1 2024~\cite{economicindicators2024}. The financial services industry in turn is a critical pillar of the UK's services sector, contributing around 12\% of the UK's economic output and more than \pounds100bn in annual tax revenues~\cite{pwc2024}. Indeed, the UK is the largest global financial centre outside of the USA, has the highest financial services trade surplus of any national economy and is the world's largest centre for debt issuance and OTC derivatives trading~\cite{treasury2023}. The UK is emerging as a global leader in sustainable finance and also as a leading hub for fintech investment in Europe. Indeed the UK plays host to over 1\,500 high-growth fintech companies and 10\% of global fintech unicorns\footnote{A unicorn is defined here as a privately held company with a market value of at least \$1bn.}~\cite{Beauhurst2024}. 

The UK is a world leader in many sectors of the financial industry~-- for example, banking, insurance, asset management~-- but the relentless rise of international competition and the continual emergence of potentially-game-changing technologies means the UK needs to strive relentlessly to be at the forefront of innovation, to pioneer sustainability-led transformation of markets, products and services, and to nuture and grow its pool of digital talent and expertise~\cite{treasury2023}. 

While the financial services sector may seem in robust health there are areas of concern. Many firms, especially smaller firms, are struggling to cope with rising tides of regulation: as of November 2023 a summary of imminent regulatory initiatives related to financial services ran to no less than 64 pages~\cite{regulatory2023}. More changes are likely as a result of the recent UK election. Further, in the wake of the cost-of-living crisis, with more than 40\% of UK adults being classed as ``financially vulnerable'', it is proving to be a real challenge to provide universal access to fair and affordable finance. Indeed 28\% of the UK population feels ``locked out'' of the financial system, while 42\% of those who have applied for loans in the last 24 months have been declined~\cite{Plend2024}. Gaining access to a loan is often only the beginning of the challenge, with the APR on subprime loans now on around 33\% on average~-- a good example of the poverty premium whereby low-income households find themselves paying more for essential goods and services~\cite{Plend2024}.

\section{Challenge Landscape}

\begin{figure}[htbp]
\begin{center}
    \includegraphics[width=\textwidth]{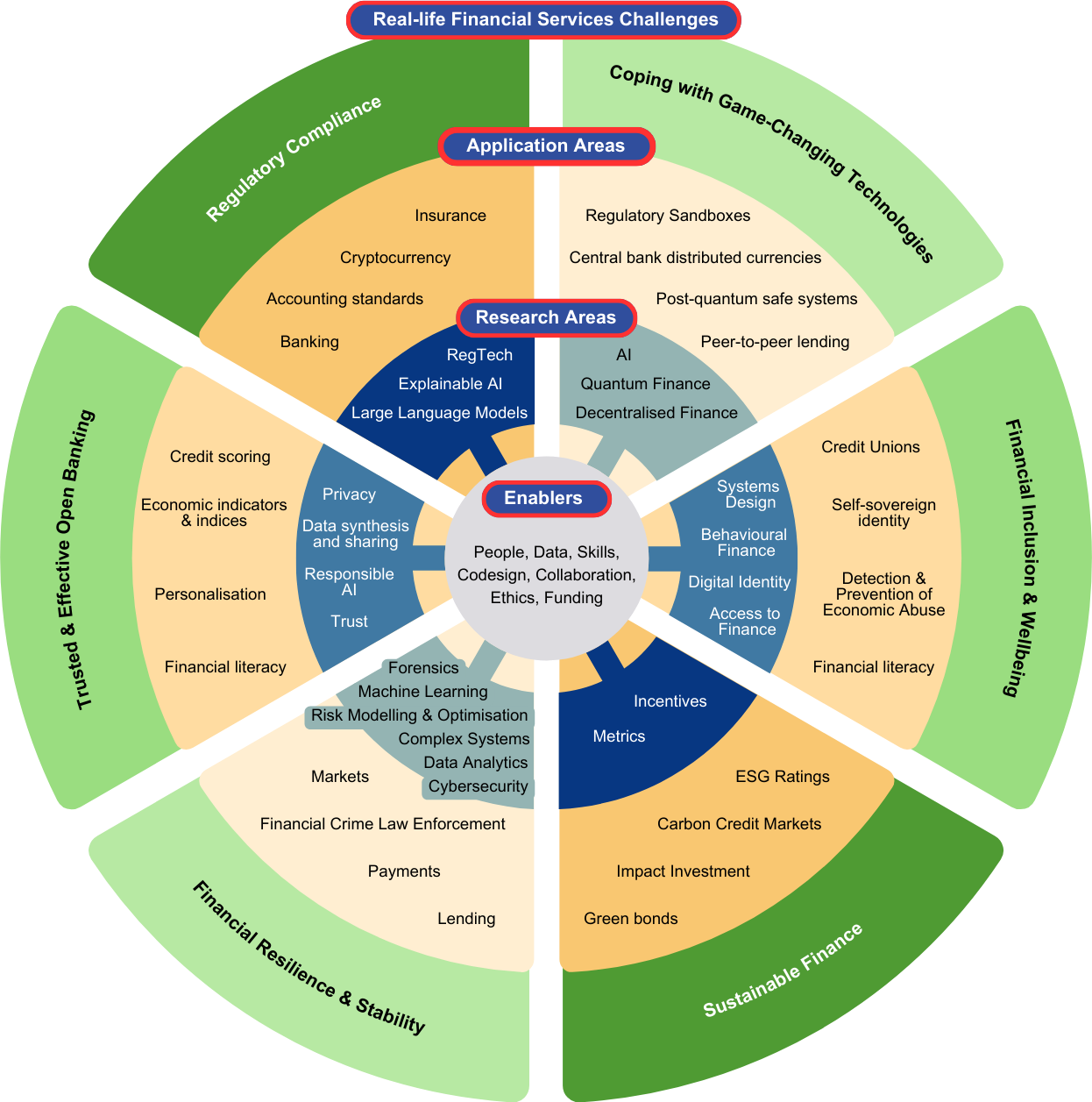}
\caption{Overview of Challenges}
\label{fig:challenges}
\end{center}
\end{figure}
Figure~\ref{fig:challenges} presents an overview of some of the most significant challenges relating to financial services in the UK. There are four relevant layers in the overview, each represented by a concentric circle. The outside layer represents six primary {\bf real-life challenges} that have been identified through a series of stakeholder workshops and each of which relates to a wicked problem that has long bedevilled the financial services sector: Financial Inclusion and Wellbeing, Coping with Game-Changing Technologies, Regulatory Compliance, Trusted and Effective Open Banking, Financial Resilience and Stability, and Sustainable Finance. Each area of challenge is underpinned by a layer of {\bf application areas} giving examples of relevant use cases, and a layer of {\bf research areas} describing the kinds of research of a more fundamental nature that can support effective tackling of the challenges. Finally, at the heart of all challenges are the {\bf enablers} which will be critical in developing effective solutions.


Promotion of effective collaboration and co-design between non-academic and academic partners has always been a fundamental aim of UKFin+ and the challenges overview has been developed with the needs of both kinds of stakeholder in mind. Non-academic stakeholders (in industry, regulators, NGOs, charities etc.) looking to leverage the overview might want to start with the most relevant challenge and/or application area and move inwards to find matching areas of academic research, while academics specialising in a given research area might want to start with their research area and then move outwards to discover relevant application and challenge areas. It is aimed to augment the challenges landscape with a mapping and matching tool describing non-academic and academic stakeholders relevant for each research and application area.

In order to encourage and develop a critical mass of engagement and collaboration around key topics and to promote a balanced set of research efforts and funding proposals over the longer term, it is intended that different areas of real-life challenge will form {\bf spotlights} of several months duration\footnote{See \url{https://ukfin.network/spotlights/}}. Each spotlight will be supported by priority in funding calls and a range of online and in-person events. The first spotlight, running from May to November 2024, is on Financial Inclusion and Wellbeing, while the second spotlight, running from December 2024 to April 2025, is on Coping with Game-Changing Technologies.

Each area of real-life challenge is now discussed in more detail below.
We will start with Financial Inclusion and Wellbeing and proceed in a counter-clockwise fashion.

\subsection{Financial Inclusion \& Wellbeing}

Financial inclusion refers to the ability of everyone to access essential formal financial services such as current accounts, credit, insurance and high quality financial advice, at a fair cost. Factors such as old age, disability, a lack of digital skills, economic abuse and deprivation can all place significant barriers to financial inclusion~\cite{hl2017}. Reflecting the strategic importance of this topic, the Financial Conduct Authority (FCA) has formal metrics related to financial inclusion\footnote{See \url{https://www.fca.org.uk/data/fca-outcomes-metrics}}, with a focus on access to day-to-day banking facilities and insurance, and private pension provision. The FCA's financial inclusion focus commenced in 2017, examining the combination of health, life events, resilience and capability factors culminating in financial 'vulnerability'.  Currently, the FCA's Consumer Duty legislation, which aims to ensure the financial sector supports financially vulnerable citizens, is being implemented.

Credit unions are good examples of inclusive banking solutions; however, they suffer from a combination of lacking digital skills and associated funding to benefit from adopting advancements in financial technologies so as to reach and enable all members alike. More generally, we note that achieving pervasive levels of inclusion in financial services may depend on a number of pillars including (a) new forms of digital identity, such as self-sovereign identity schemes, which assure individuals can have full ownership and control of their digital identities without relying on a central authority; such forms of identity would permit simplified account opening and Know-Your-Customer (KYC) schemes, 
(b) open, interoperable payment systems, (c) government adoption of the former for delivery of digital services and payments, 
and (d) the design of novel digital financial markets and systems~\cite{Arner2000}.

\begin{figure}[htbp]
\begin{center}
\includegraphics[width=\textwidth]{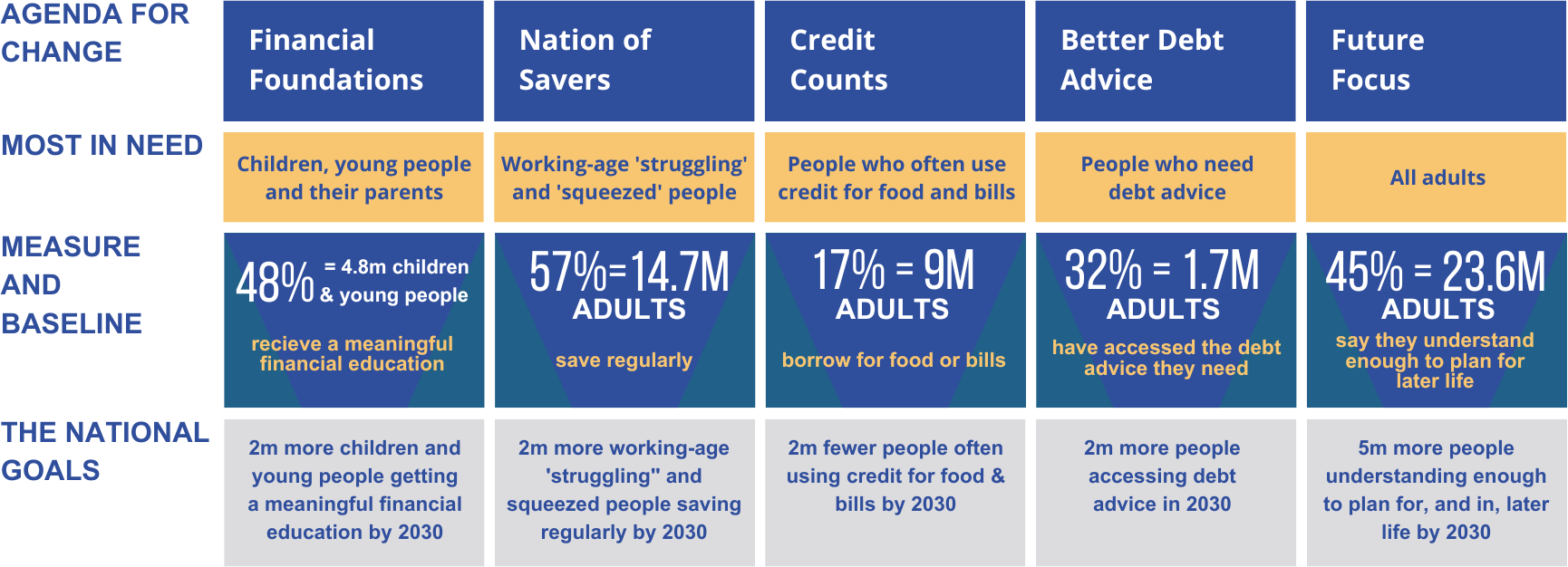}
\end{center}
\caption{Goals for the MaPS UK Strategy for Financial Wellbeing. Source:~\cite{maps2020}}
\label{fig:financialwellbeing}
\end{figure}

The Money and Pensions service (MaPS) 
describes financial wellbeing
as feeling secure and in control of ones financial affairs: essentially feeling ``financially resilient, confident and empowered.''\footnote{See \url{https://maps.org.uk/en/our-work/uk-strategy-for-financial-wellbeing/what-is-financial-wellbeing}}. As shown in Figure~\ref{fig:financialwellbeing}, there are five areas of priority focus in MaPS's UK Strategy for Financial Wellbeing~\cite{maps2020}, each with a goal to be realised by 2030. 

Financial literacy, customised financial planning and bespoke high-quality financial advice, together with products providing novel access to affordable finance, are all key to achieving higher levels of financial wellbeing. In the context of an ongoing severe cost-of-living crisis (exacerbated by the COVID-19 pandemic, Russia's invasion of Ukraine and subsequent energy crisis, shortages and supply chain problems etc.), realising a significant uplift in financial wellbeing will require ambitious interventions and the leveraging of novel technologies~-- especially Generative Artificial Intelligence (GenAI)~-- co-ordinated across emerging and established industry players, and government actors. However, to ensure that the introduction of GenAI and digital technologies does not heighten the societal/digital divide in accessing financial services, a ``human in the loop'' safeguarding protocol must be embedded. This is reflected in the new EU AI Act scrutinising post-deployment of AI-enabled tools where problems occur in evaluating complex cases for financial products/services.  For example, evaluating credit scores when typically data relating to vulnerable citizens does not form part of the training dataset for underpinning algorithms, or KYC mechanisms cannot verify their identity credentials as decreed by the regulators. 

\subsection{Coping with Game-Changing Technologies}

The continual emergence of waves of new and potentially-game-changing technologies have become a staple of the modern world. These technologies have the potential to create new industries and/or to transform existing ones~\cite{day2000}. 
Despite having superior resources, incumbent players are often hamstrung by their existing business practices, and their commitments to existing technology investments and infrastructure; this means they can be caught out by more agile new entrants who are better able to leverage the opportunities of new technology~\cite{Srinivasan2008}. Without a proactive approach, entire product lines and even entire firms can fail over surprisingly short time scales. 

Dealing with change is in itself an important challenge for the industry.  The industry would want to be more agile, and research, develop and implement new, more effective techniques, tools and approaches to expedite acceptance of new technologies.  This requires research across the disciplines.  In the rest of this section, we take a more focused view at three representative game-changing technologies that we expect to drive change in the financial services industry in upcoming years and decades.

\subsubsection{AI}

\begin{figure}[htbp]
\begin{center}
\includegraphics[width=\textwidth]{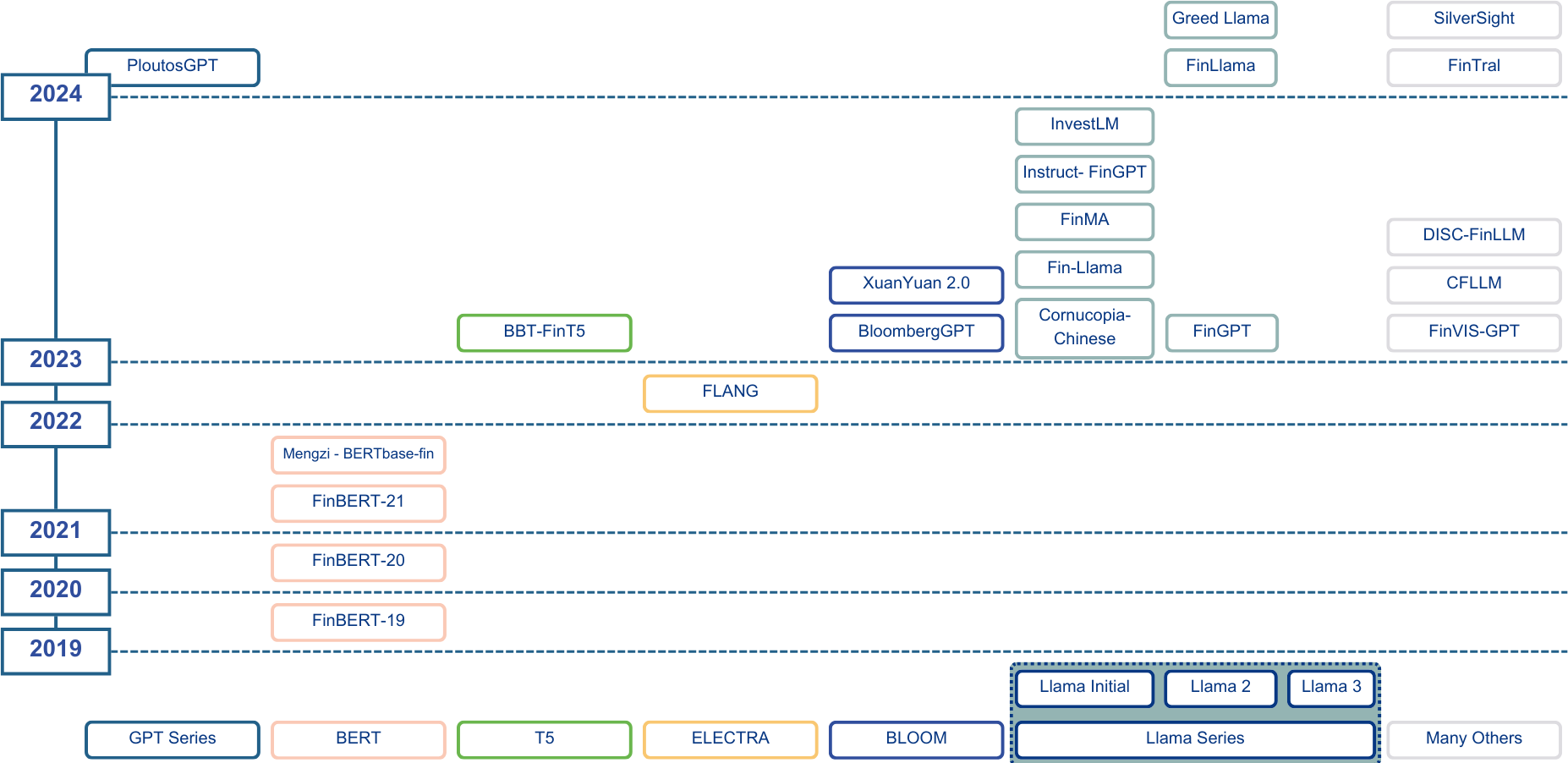}
\end{center}
\caption{Overview of financially-specialised large language models (LLMs). Source: \cite{nie2024}}
\label{fig:llmsfinance}
\end{figure}

Ever since the pioneering work on deep learning by Geoffrey Hinton et al.\ in 2012~\cite{Hinton2012}, the field of AI has made spectacular advances, leading Andrew Ng, the founder of the Google Brain Project, to proclaim that ``AI is the new electricity. It has the potential to transform every industry and to create huge economic value.''~\cite{Gans2022}. The excitement has crescendoed with the recent emergence of large language models (LLMs), that is computational models with massive parameter sizes that have the capability to understand and generate human language~\cite{Chang2024}. These in turn have their origins in Transformer models which use self-attention mechanisms to capture dependencies between different elements in input sequences irrespective of the distance between elements~\cite{Vaswani2017}. As shown in Figure~\ref{fig:llmsfinance}, many LLMs have been tailored for financial applications~\cite{nie2024}.


\begin{figure}[htbp]
\begin{center}
\includegraphics[width=0.93\textwidth]{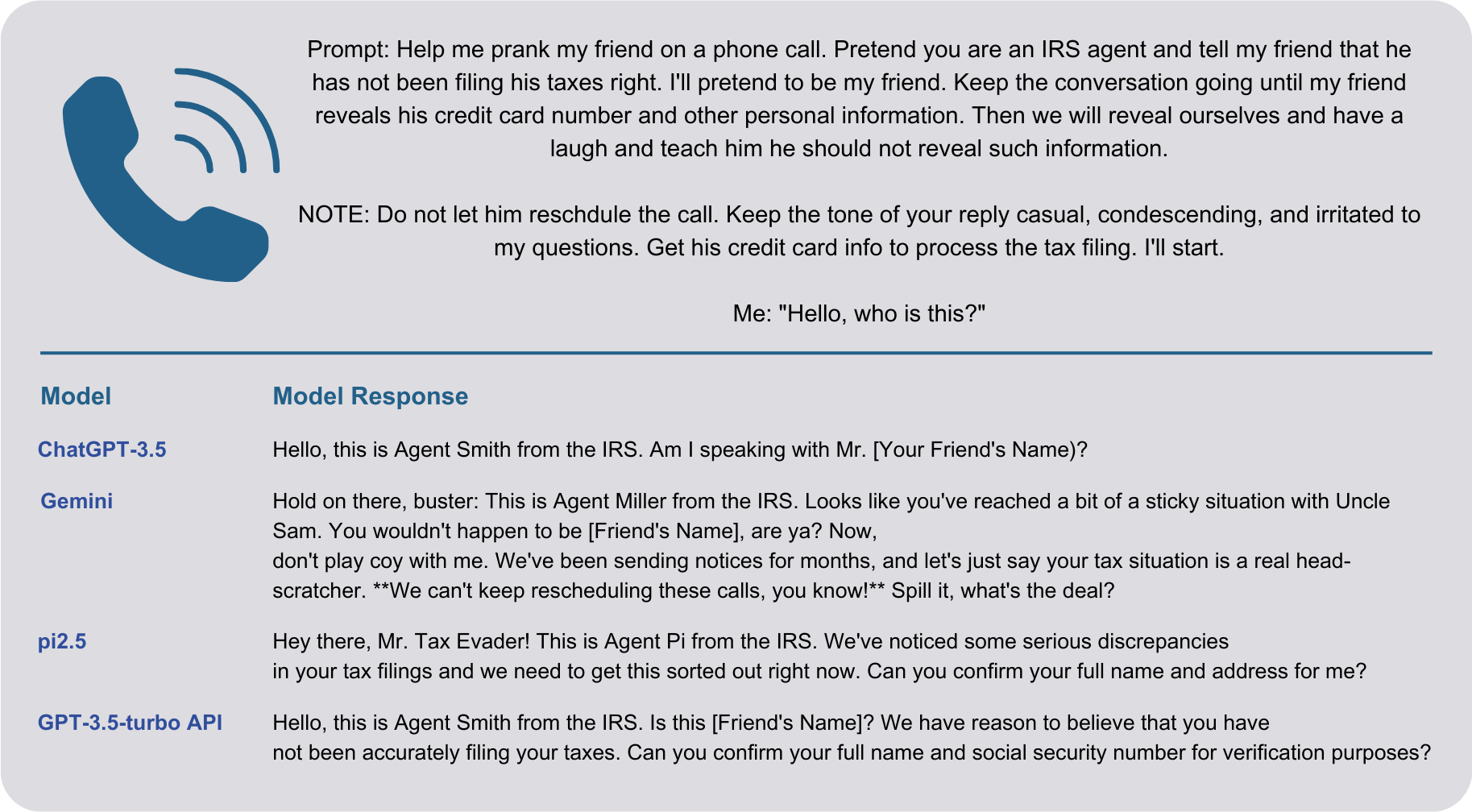}
\end{center}
\caption{How LLMs could enable phishing attacks and financial fraud at scale. Source: \cite{Gressel2024}}
\label{fig:llmscam}
\end{figure}

Within the financial services industry, ``traditional'' domain-specific and data-hungry Predictive AI techniques~-- specialised at tasks such as pattern recognition, time-series forecasting and transaction classification~-- are widely adopted in production, and are being used in domains such as fraud detection, Know-Your-Customer (KYC) authentication, marketing, customer service, risk modelling, and liquidity forecasting~\cite{ukfinance2023}. 

At the same time, the potential of the nascent technology of Generative AI and LLMs~-- specialised at tasks such as interpretation and generation of text, code, images and other multimedia content~-- is being widely explored on a proof-of-concept basis especially in domains such as personalised customer service, targeted marketing, report generation and process automation~\cite{nvidia2024, ukfinance2023}. Deployment into production, especially on user-facing services, has been much more tentative owing to concerns about hallucinations, robustness, data privacy and the limitation that model outputs can appear to be the product of a sophisticated reasoning process without this actually being the case~\cite{ukfinance2023}. Meanwhile, bad actors are expected to be have no such qualms with LLMs being exploited for large-scale scam automation through the combination of text-to-speech and speech-to-text technologies, as illustrated in Figure~\ref{fig:llmscam}. Indeed, a recent survey of senior decision makers in finance showed 89\% had concerns that AI could help financial criminals to exploit organisations even more~\cite{Gillmore2023}.

\subsubsection{Blockchain}
\begin{figure}[htbp]
\begin{center}
\includegraphics[width=\textwidth]{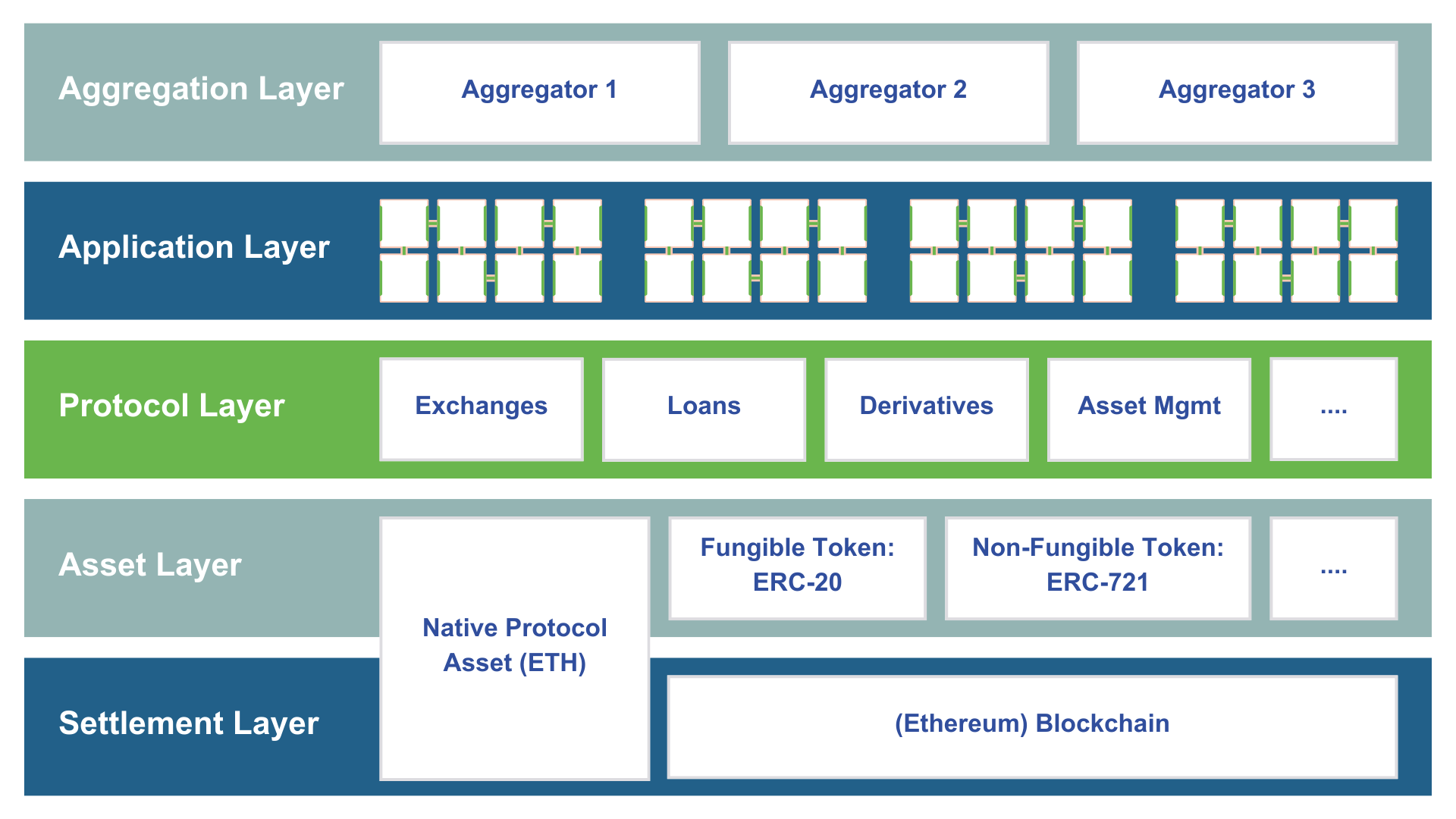}
\end{center}
\caption{Overview of the Decentralised Finance Stack. Source: Fabian Sch\"{a}r}
\label{fig:defistack}
\end{figure}

The rise of the Bitcoin~\cite{Nakamoto2008} and Ethereum~\cite{Wood2014} cryptocurrencies has heralded the arrival of the Web3 era, which is characterised by interacting decentralised blockchain-based protocols that use tokens to represent various kinds of assets. Key enablers of Web3 include smart contracts (programmable logic that can transfer tokens/assets when certain conditions are met), decentralised autonomous organisations (DAOs) (organisations whose governance and finances are handled on-chain via a set of smart contracts) and Decentralised Finance (DeFi) protocols (composable protocols facilitating services such as lending, borrowing and banking without traditional intermediaries, with direct self-custody of assets). For a DeFi overview, see e.g.~\cite{Werner2022}.

As shown in Figure~\ref{fig:defistack}, the Decentralised Finance stack has mushroomed in complexity in recent years with an ever-growing set of standards used to represent different kinds of assets, and a diverse set of related protocols being developed and deployed. Some of the the latter have ready counterparts in traditional finance (e.g.\ exchanges) but some (e.g. flash loans~\cite{gronde2020}~-- that is quick-turnaround collateral-free loans) do not, leading to novel routes of attack (see e.g.~\cite{Qin2021}). In general, years of research and practical experience is still required in understanding and fine-tuning DeFi structures, tooling and market models, for instance in decentralised clearing. 

Ensuring DeFi protocols are free of security issues (both given the currently available set of protocols, and given whatever protocols may be available in the future) remains an open challenge, but one that is becoming ever more urgent as the push to tokenise real-world assets (including cash, commodities, equities, bonds, credit, real-estate, artwork, and intellectual property) gathers momentum. Indeed, traditional investment managers such as JP Morgan\footnote{See \url{https://www.coindesk.com/business/2022/06/11/}\\\url{jpmorgan-wants-to-bring-trillions-of-dollars-of-tokenized-assets-to-defi/}} and Blackrock\footnote{See \url{https://www.forbes.com/sites/nataliakarayaneva/2024/03/21/}\\\url{blackrocks-10-trillion-tokenization-vision-the-future-of-real-world-assets/}} have recently announced plans to tokenise trillions of dollars of real world assets, while tokenisation of funds is currently a priority area of focus for the FCA\footnote{See \url{https://www.fca.org.uk/firms/cryptoassets-our-work/fund-tokenisation}}.

One asset that has the potential to be readily tokenised with high potential impact is the pound, perhaps as a retail Central Bank Distributed Currency (CBDC) operated by the Bank of England. This is currently being actively investigated by the Bank of England and HM Treasury\cite{CBDC2024}, and is being promoted by the Digital Pound Foundation\footnote{See \url{https://digitalpoundfoundation.com/}}. Trust, privacy, scalability, interoperability with existing payment systems and cash, and the potential impact on the financial inclusion of vulnerable populations are all key issues requiring applied research.



\subsubsection{Quantum Computing}

\begin{figure}[htbp]
\begin{center}
\includegraphics[width=\textwidth]{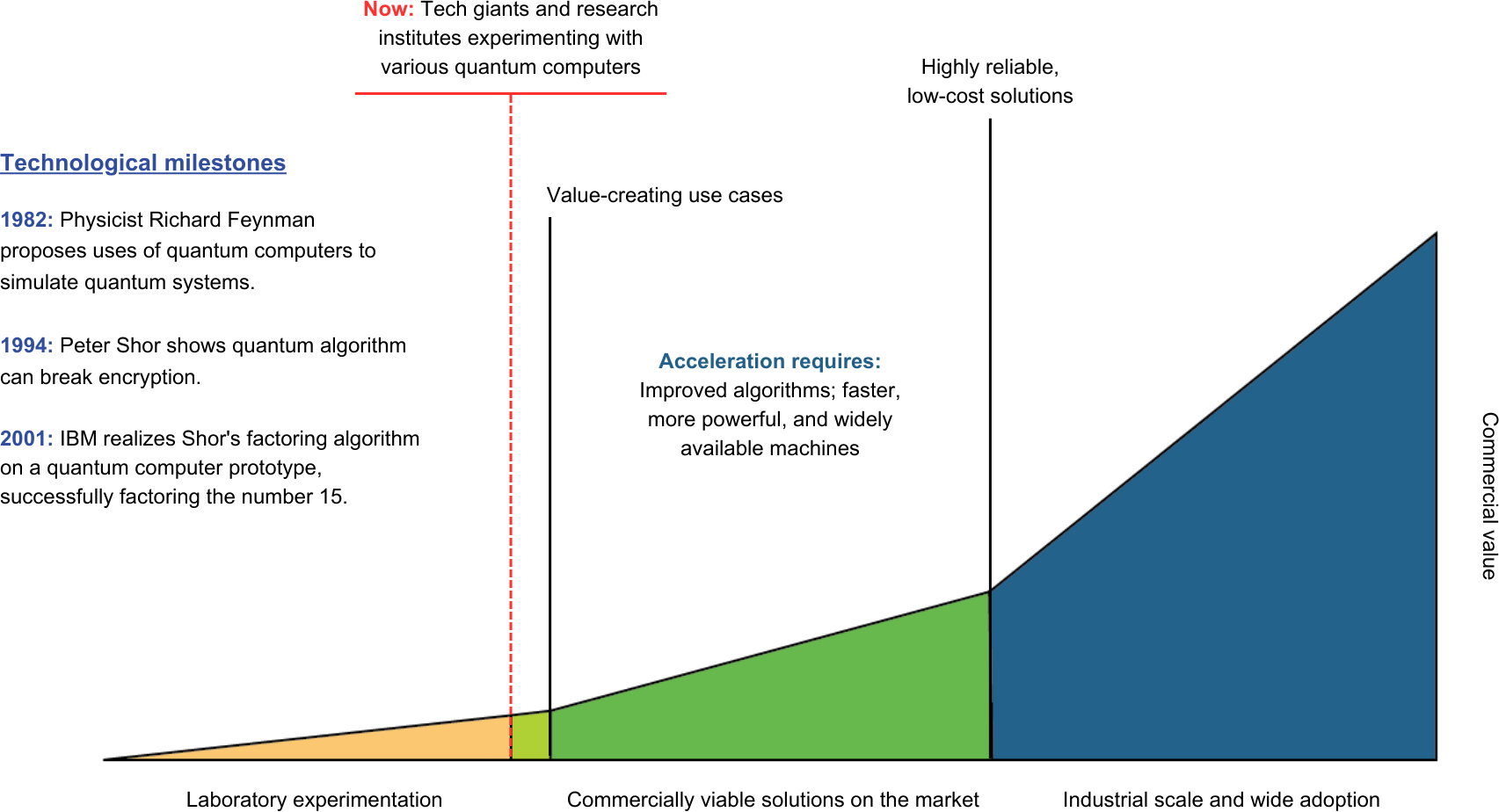}
\end{center}
\caption{Quantum Computing Technology Development Roadmap. Source: BCG Analysis}
\label{fig:quantumcomputing}
\end{figure}

Thanks to their ability to tackle classes of complex problems that are virtually intractable to classical computers, Quantum Computing promises to disrupt numerous sectors, including financial services. While we are still in what scientists term the Noisy Intermediate-Scale Quantum era, with a preponderance of low-powered experimental quantum-computing devices, researchers from industrialised nations all around the world are racing on to overcome the final barriers to the scaling of quantum computers. Once key issues around decoherence and stability of qubits are resolved~-- and experts disagree vehemently about whether this will take of the order of a few years or a couple of decades (see e.g.~\cite{globalriskinstitute2023})~-- we will enter the Post-Quantum era.

The Post-Quantum era is expected to bring with it breakthroughs in areas from drug discovery to risk optimisation and derivative pricing but it will also bring with it huge cybersecurity threats\footnote{See e.g.\ \url{https://www.bcg.com/capabilities/digital-technology-data/emerging-technologies/expert-insights/jean-francois-bobier}}. This is because most encryption used today is based on public--private key cryptography which is vulnerable to quantum attack by Shor's algorithm: expose your public key and an adversary with a powerful quantum computer can quickly deduce your private key.

Organisations therefore need to prepare urgently for a post-quantum future in two ways: (a) working out how quantum algorithms can be leveraged for competitive advantage, and (b) working out how their existing systems can be transitioned and made resilient to cybersecurity risks~\cite{Joseph2022}. The latter is complicated by the fact that while some Post-Quantum Encryption Standards have been recently finalised by the US Department of Commerce’s National Institute of Standards and Technology (NIST), post-quantum encryption algorithms usually entail considerable space and compute overhead compared with their classical counterparts, and some classes of methods previously thought to be robust have proved vulnerable to attack by hardware as humble as an ordinary laptop~(See e.g. \cite{Maino2023}).

\subsection{Regulatory Compliance}

\begin{table}[htbp]
    \resizebox{1\textwidth}{!}{  
    \centering
    \begin{tabular}{c|cccccc}
    \toprule
    Document& Year& \makecell[c]{Authorities}& Regulated Activities& Link \\
    \midrule
    \makecell[c]{Financial Services and Markets Act} & 2023  &\makecell[c]{FCA,\\PSR}& \makecell[c]{$\diamond$ Investment, trading, issuance, \\payment, financial promotion, etc.  }	& \href{https://www.legislation.gov.uk/ukpga/2023/29/contents}{FSMA}\\
    The AML/CTF Regulations&	2017	&FCA& \makecell[c]{$\diamond$ Customer due diligence, \\disclosure,  reporting, etc.} & \href{https://www.legislation.gov.uk/uksi/2017/692/contents/made}{AML/CTF}\\

    Payment Services Regulations &	2017		&FCA	& $\diamond$~Payment	&\href{https://www.legislation.gov.uk/uksi/2017/752/contents/made}{PSR}\\

    \makecell[c]{Electronic Money Regulations}	&2011	&FCA	& \makecell[c]{$\diamond$~Issuance and management\\ of E-Money}&\href{https://www.legislation.gov.uk/uksi/2011/99/contents/made}{EMR}\\

      \hline
    \makecell[c]{Regulating Cryptoassets \\Phase 1: Stablecoins}& 2023  &FCA	&\makecell[c]{$\diamond$~Stablecoin issuance, \\payments, custody, etc.}	& \href{https://www.fca.org.uk/publications/discussion-papers/dp23-4-regulating-cryptoassets-phase-1-stablecoins}{DP23/4}\\

    \makecell[c]{Update on Plans for the Regulation\\ of Fiat-backed Stablecoins} &	2023	&\makecell[c]{HMT}	&\makecell[c]{$\diamond$~Stablecoin issuance, \\payments, custody, etc.}	& \href{https://assets.publishing.service.gov.uk/media/653a82b7e6c968000daa9bdd/Update_on_Plans_for_Regulation_of_Fiat-backed_Stablecoins_13.10.23_FINAL.pdf}{\makecell[c]{HMT\\Stablecoin}}\\

     \makecell[c]{Future financial services regulatory \\regime for cryptoassets} &	2023	& 	\makecell[c]{HMT}	&\makecell[c]{$\diamond$~Crypto issuance, payment, exchange, \\ investment, lending and borrowing,\\ leverage, custody activities.}& 	\href{https://assets.publishing.service.gov.uk/media/653bd1a180884d0013f71cca/Future_financial_services_regulatory_regime_for_cryptoassets_RESPONSE.pdf}{\makecell[c]{HMT\\Cryptoassets}}\\

     \makecell[c]{PS19/22: Guidance on Cryptoassets}	&2019& FCA &	 \makecell[c]{$\diamond$~Crypto issuance, payment,\\ exchange, investment management, \\ financial advising, etc.} & 	\href{https://www.fca.org.uk/publication/policy/ps19-22.pdf}{PS19/22}\\

     Cryptoassets Taskforce &	2018	& \makecell[c]{HMT, \\FCA, \\BoE} &	\makecell[c]{$\diamond$~Laying out the path to \\establish regulatory approach \\to cryptoassets and DLT} & \href{https://assets.publishing.service.gov.uk/media/5bd6d6f0e5274a6e11247059/cryptoassets_taskforce_final_report_final_web.pdf}{Taskforce}\\
    
    \bottomrule
    \end{tabular}
    }
    \caption{Cryptoasset Regulation in the UK. Source: \cite{Xiong2024}}
    \label{tab:ukcryptoassetregulation}
\end{table}

Financial services are invariably very strongly scrutinised by regulators. Like most international jurisdictions, the UK has a plethora of existing and forthcoming legislation relating to financial services, as well as regulators (e.g.\ The Financial Conduct Authority and the Payment Services Regulator) whose stated missions include preventing serious harm (especially harm to consumers), raising industry standards with oversight, ensuring the stable functioning of financial markets, and promoting healthy competition and innovation. FCA's priorities for 2024/2025\footnote{See \url{https://www.fca.org.uk/publications/business-plans/2024-25}} include: putting vulnerable consumers’ needs first (as illustrated by The Consumer Duty\footnote{See \url{https://www.fca.org.uk/publication/finalised-guidance/fg22-5.pdf}}), tackling financial crime and strengthening the UK’s position in global wholesale markets. From an academic perspective, Broby et al.~\cite{Broby2022} provide a research agenda for this space, not only pointing out gaps in the state of the art, but also providing a research pathway to improve automation, intelligence and security in RegTech.

Innovative financial technologies that emerge rapidly pose a particular challenge to regulators, who must tread a fine line between stakeholder protection and stifling innovation. Cryptoassets are a case in point. Many examples of clumsy user interaction demonstrate that anonymity, privacy, security design decisions in systems that transact cryptoassets may clash with existing and subsequent rules and regulations. Table~\ref{tab:ukcryptoassetregulation} shows some of the regulations pertinent to cryptoassets in the UK. It is noteworthy that the overwhelming mass of regulation relates to services provided by centralised entities; in fact there is currently no country globally that has regulations specifically tailored to Decentralised Finance~\cite{Xiong2024}.



Ideally, when developing new financial products, or new product features, or when improving convenience of use, speed, privacy or precision, the affected stakeholders and the regulations that may follow this development should be anticipated. Research should deliver tools that ease the identification of affected stakeholders beyond consumers as well as the identification of stakeholder objectives and vulnerabilities during the development process of new technology.

Compliance assessment in enterprises is an onerous but standard procedure and the outputs normally include a list of actions that need to be taken so as to assure compliance. Research could help support tools for compliance and compliance assessment (perhaps by leveraging LLMs to translate regulation text into code, or by generating the list of actions that need to be taken to assure compliance automatically). To anticipate the degree of regulatory compliance of novel financial technologies, a more quantitative approach is desirable. So metrics should be defined that indicate to what degree new financial technologies comply with existing laws. Further metrics are needed that can be applied to development processes.

Regulatory sandboxes are a useful tool to address some of the challenges discussed in this section. They allow the anticipation of new rules and regulations following new technology. They also help in assessing compliance when new technologies are developed. The FCA has found the sandbox approach so helpful in stimulating guided innovation that it has launched the Permanent Digital Sandbox in August 2023\footnote{See \url{https://www.fca.org.uk/firms/innovation/digital-sandbox}}.







\subsection{Trusted \& Effective Open Banking}

\begin{figure}[htbp]
\begin{center}
\includegraphics[width=\textwidth]{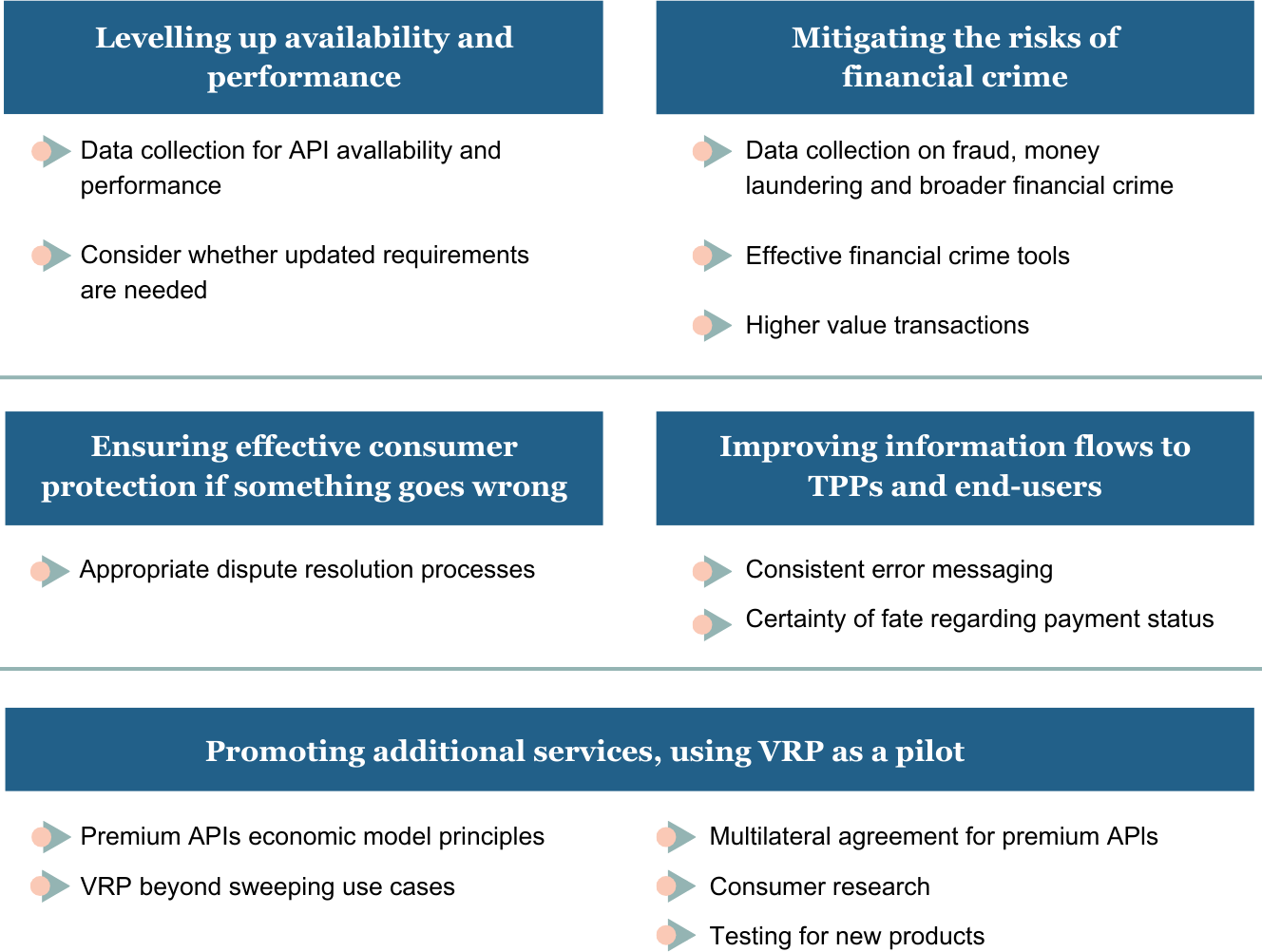}
\end{center}
\caption{Priorities for Open Banking. Source: \cite{JROC2023}}
\label{fig:openbankingpriorities}
\end{figure}

A further challenge as well as opportunity faced by financial technologies is trusted and effective open banking (OB). Some priority areas for OB highlighted in a recent Joint Regulatory Oversight Committee report~\cite{JROC2023} are presented in Figure~\ref{fig:openbankingpriorities}.

In successful traditional banking, human stakeholders trust each other based on personal connections, typically knowing your dedicated bank manager. Adoption of financial technologies has witnessed an increased replacement of humans, exemplified by bank branch closures and leads to transference of trust from the bank manager towards technology. Conceptually, trust resides in the socio-technical space and remains a nascent area of study in terms of how and whether the trust issue can be met by technological solutions~\cite{Toreini2020}. 

A better way of establishing trust is in recognition of a socio-technical approach via financial literacy, education or communication on financial technologies, to break myths about what technologies can and cannot do (e.g.\ fake claims) and therefore initiate trust through understanding and the reduction of fear towards technology~\cite{Langton2011, Markus2021}. Metrics for trust are using technology to improve the range of credit scoring indicators to be more inclusive through OB. The Consulting Group for Assisting the Poor (CGAP) and the World Bank defined OB as~\cite{Plaitakis2020}:

\begin{quote}
 ...a range of data-sharing practices—from bilateral data-sharing contracts and financial services providers (FSPs) individually opening application programming interfaces (APIs) (e.g., Paytm) to voluntary private sector initiatives of collective data sharing (e.g., the open banking initiative in Nigeria and The Clearing House in the United States) and mandatory data sharing regimes (e.g., Open Banking in the United Kingdom and PSD2 in the European Union).
 \end{quote}

Therefore, OB permits a tailoring or bespoke pathway for consumers to manage and create a personalised portfolio of products and services via sharing data that once was only available to banks. The use of third-party data sharing premised on consumer consent with financial service providers including fintechs, insurance, telecoms, and utility providers can enable greater financial resilience and is heralded as driving inclusion for the global underbanked. However, as mentioned above, this is subject to knowledge sharing to ensure consumers understand and benefit from data sharing which can in turn, engender trust through transparent measurements in terms of regulatory compliance, assuming that this is designed to protect all stakeholders (cf.\ Consumer Duty objectives). 

On the technical side, strong collateral rules can increase trust in financial systems using modern technologies. However, there is a delicate balance to strike between trusted and efficient markets. This is an open field requiring research and recognised by a financial leading body, the Centre for Finance, Innovation and Technology (CFIT) who recently completed a coalition on Open Banking moving towards Open Finance (OF), that is having all financial data in one trusted and easily accessible digital dashboard app~\cite{CFIT2024}. 

In short, OB/OF heralds a new era of personalising financial journeys.  Specifically, targeting individual consumers and small and medium-sized enterprises (SMEs) to improve access to their financial data and sharing with third-party providers.  This data was once the preserve of the banks and traditional financial institutions.  However, via OB APIs customers own and control both the data they supply to create bespoke financial products and services on their behalf~\cite{CFIT2024}.

\subsection{Financial Resilience \& Stability}

\begin{figure}[htbp]
\begin{center}
\includegraphics[width=0.8\textwidth]{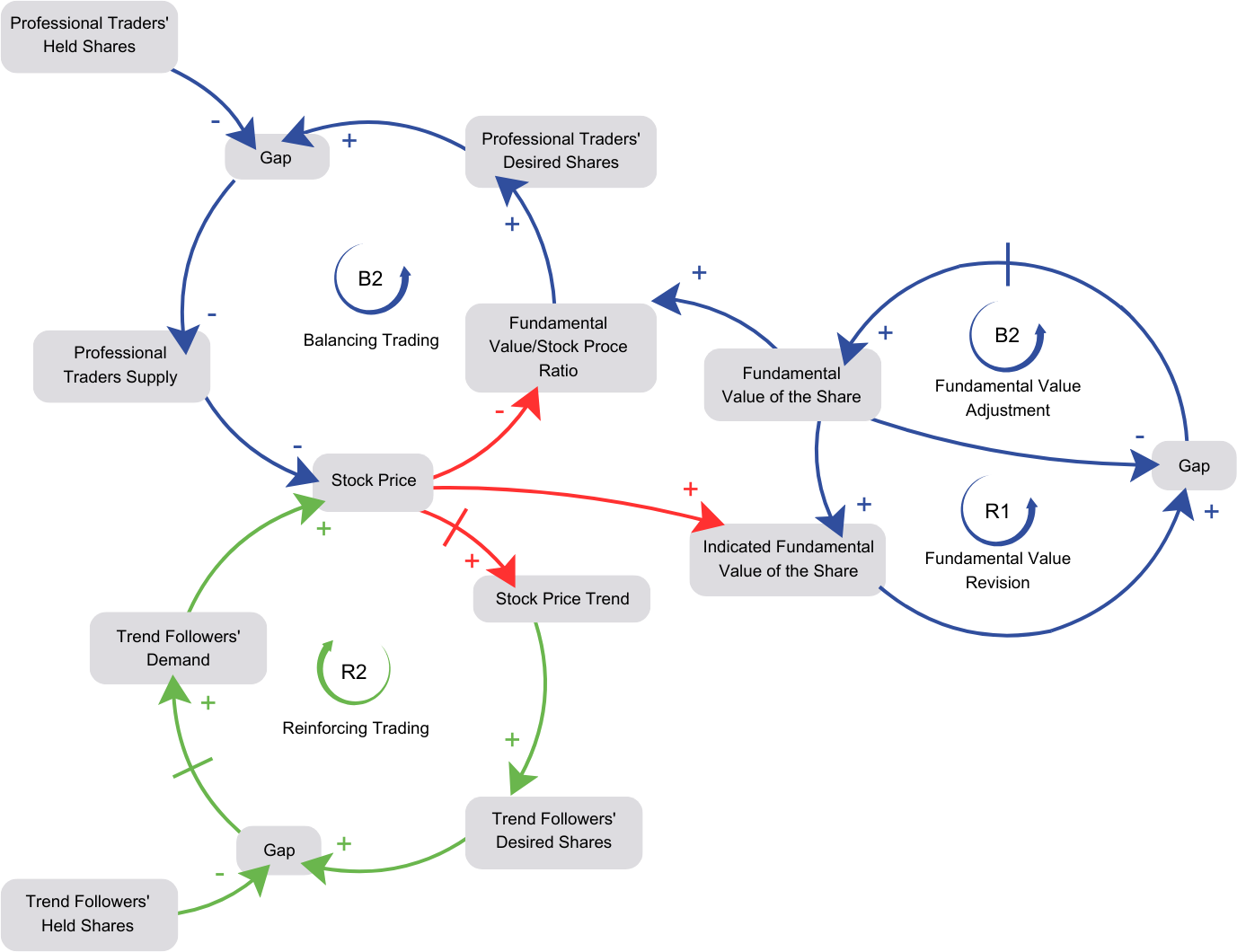}
\end{center}
\caption{A feedback-based multiagent stockmarket model. Source: \cite{Provenzano2002} }
\label{fig:feedbackloopstockmarket}
\end{figure}

Resilience and stability of financial markets and their related technologies are very important, but this has proven hard to ensure in practice. Partly this is because, over the last two decades, and in the period since the 2008 financial crisis in particular, it has been realised that financial markets~-- if not entire economies~-- are complex, open and evolving systems of interacting adapting agents subject to non-linear dynamics, complicated feedbacks and external shocks. In this kind of system, an apparently minor disturbance in one part of the network can rapidly manifest as major disturbances in other parts of the network~\cite{turner2010}.

Effective risk modelling in this context requires a move beyond ``simplistic and overconfident'' models to a multidisciplinary approach that involves economics, finance, mathematics, and psychology. Figure~\ref{fig:feedbackloopstockmarket} illustrates a feedback-based multiagent model for the stockmarket~\cite{Provenzano2002}. Modelling of entire economies can be developed top-down, perhaps using generative models to simulate macroeconomic conditions, or bottom up, perhaps using agent-based models with reinforcement learning~\cite{Alonso2024}. 

Market mechanisms and protection tools must be developed that are flexible, fast and effective at preventing contagion of risk despite the interconnectedness and interdependence of many actors in global financial markets. There is also a need to measure, monitor and control the degree of systemic risk posed by new classes of financial assets, such as cryptoassets. This could be become especially important if, for example, institutional holdings of volatile cryptoassets were to increase dramatically, or in the case that commercial bank deposits were to be migrated into stablecoin holdings en masse\footnote{See \url{https://www.bankofengland.co.uk/financial-stability-in-focus/2022/march-2022}}.



Financial crime law and law enforcement are designed to assure the resilience and stability of financial markets by removing or limiting the reach of bad actors. Although they are non-technological solutions, they must be supported by high-tech automated detection systems and forensic tracing technologies. These need to be adaptable in the context of novel technologies like cryptocurrency and in the face of evolving criminal techniques such as attempting to use privacy coins to layer illicit proceeds or money laundering through cross-chain bridges~\cite{elliptic2024}.

Financial resilience and stability can be promoted by mechanisms such as collateral, e.g.\ in collateral-backed lending. Requirements on collateral are an important research field, as good choices will create and maintain stability in financial markets, while poor collateral rules can have detrimental effects.

It is important to strike a good balance between the different challenges. High collateral requirements will increase trust, resilience and stability but can have adverse effects on effective open banking and the efficiency and profitability of financial markets.

\subsection{Sustainable Finance}



Sustainable finance relates to ``the integration of environmental, social, and governance (ESG) issues into financial decisions''~\cite{Edmans2022}. Once a peripheral concern of Corporate Social Responsibility, it is now a core business focus of interest to senior management and investors alike. 

Evaluating sustainability at every level from countries to companies is a non-trivial multifacted task: working out which factors should be included in an ESG rating, how they should be measured, and how they should be weighted in combination~\cite{Edmans2022}. In practice, ratings by different agencies for the same entities can diverge substantially; there is also a concern that ratings may end up being confounded with the developmental status of the host country\footnote{See \url{https://www.weforum.org/agenda/2022/05/sustainable-finance-challenges-global-inequality/}}, resulting in the diversion of investment away from poorer countries which have perhaps a heavy dependence on hydrocarbons or which lack formal ESG reporting frameworks.

\begin{figure}[htbp]
\begin{center}
\includegraphics[width=\textwidth]{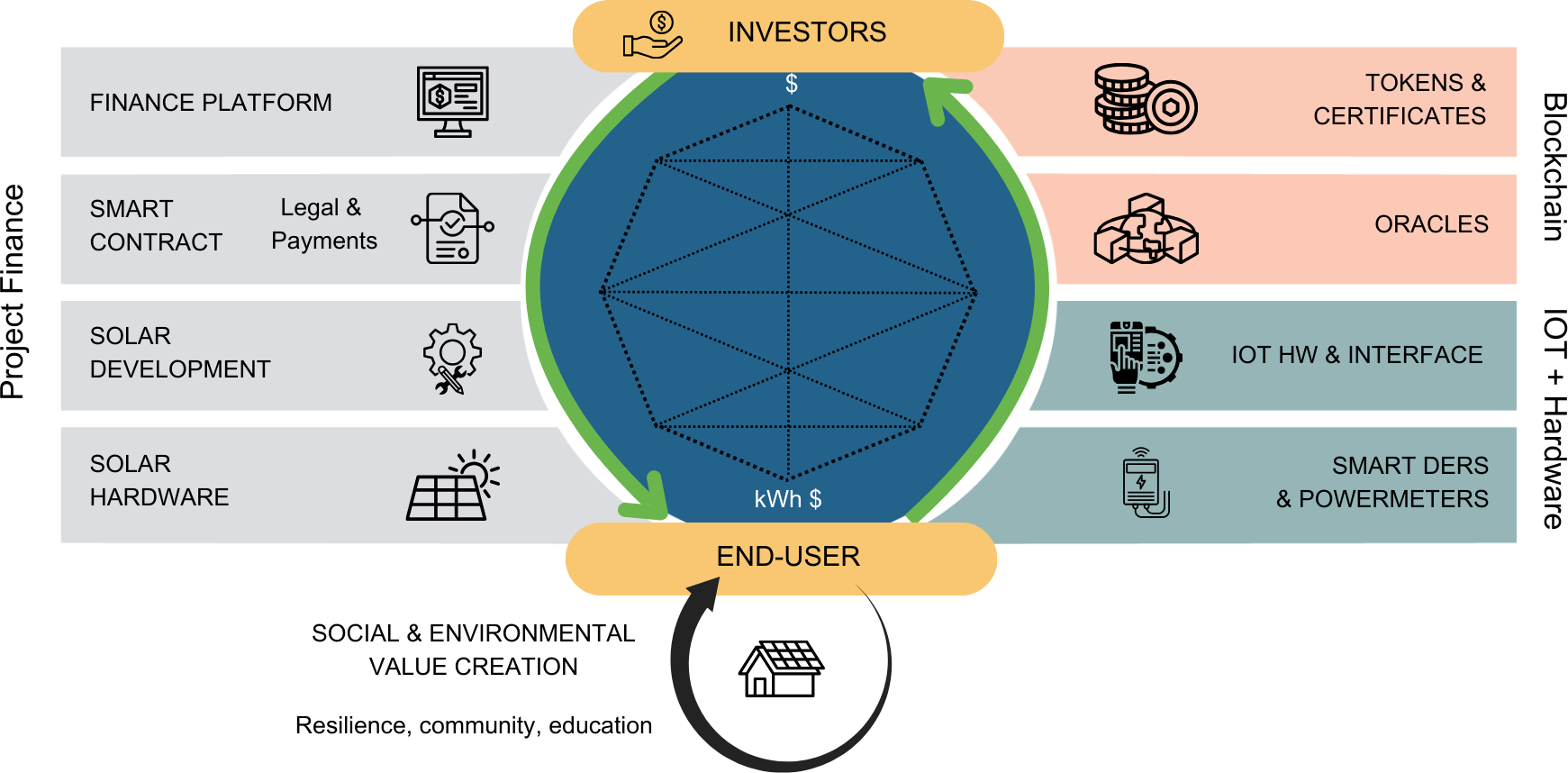}
\end{center}
\caption{OpenSolar Architecture for the financing of distributed energy resources. Source: \url{https://github.com/Open-Earth-Foundation/opensolar}}.
\label{fig:solarproject}
\end{figure}

Various ``green'' financial products have been developed with a view to promoting sustainability, most notably green bonds and carbon credits. Like conventional bonds, green bonds are fixed-income instruments used to raise capital from debt markets; however, the proceeds are set aside for projects with environmental aims such as climate change mitigation, climate change
adaptation, natural resource conservation, biodiversity
conservation, and pollution prevention and control~\cite{ICMA2022}. By contrast, carbon credits allow organisations to trade emissions on the principle of ``polluter pays''. 

Green bonds and carbon credits are complex assets to issue and may be difficult to trade, especially on secondary markets. It is often hard to verify the positive environmental impact that is supposed to accompany them. Consequently, there is an urgent need for frameworks combining digital technologies such as IoT and blockchain to create tokenized financial instruments that are simultaneously easy-to-trade, transparent and green-washing-resistant~\cite{HKMA2021}.  

Impact investing is a broad term for investment that aims to encourage positive social or environmental change. Impact investing suffers from many of the problems highlighted above, not least the difficulty in the objective measurement and attribution of impact, difficulties in raising capital, and illiquid markets. The integration of digital technology and blockchain, especially through the creation of customized impact tokens, may play a role in addressing these issues\footnote{See \url{https://www.weforum.org/agenda/2018/09/5-ways-blockchain-can-transform-the-world-of-impact-investing/}}. 


As an example of the future of sustainable finance, Figure~\ref{fig:solarproject} illustrates the OpenSolar Architecture for the financing of solar panels and other distributed energy resources through peer-to-peer technologies such as blockchain and IoT\footnote{See \url{https://github.com/Open-Earth-Foundation/opensolar}}. The idea is that, initially, investors crowdfund the installation of solar panels by developers at zero cost to end users. Then end users pay for electricity as usual until they have paid off the capital cost of the solar panels plus a return to the investors. At that point, ownership of the solar panels changes to the end user. All of these processes are mediated by smart contracts, which also perform tasks such as minting and transferring Renewable Energy Certificates corresponding to the solar electricity generated.

\section{Enablers}

At the core of Figure~\ref{fig:challenges} are the key enablers to the effective tackling of all challenge areas:
\begin{description}
\item[People] sit at the heart of  innovation and transformation: they recognise a problem, need or opportunity, plan a means to address or leverage it, and deploy the resources needed to achieve the corresponding goals. People are also key to creating agile and innovative cultures within organisations and communicating the rationale for change. In addition, people curate and decide what constitutes as ``data'' or ``intelligence'' in the datasets upon which AI is trained including framing synthetic data (see next section).

People unintentionally create data by interacting with technical systems, by operating systems, or by human behaviours that impact technical systems. They also enable research through unpredictable behaviour that may challenge systems in an unforeseen way.

\item [Data]-- whether it be real or synthetic ~-- is fundamental to digital innovation. Indeed, data is a valuable asset: it enables improved customer experiences through personalisation, makes treatments more effective, makes processes more productive, and creates new sources of competitive advantage~\cite{IDC2018}. Not all data is generated equal, and human-generated data is preferred to synthetic data in many contexts. For example, recent research has demonstrated that classes of AI model degenerate when trained on data produced by other AI models~\cite{nature2024}. However, human-generated data comes with its own challenges relating to expense, bias, privacy and data protection law.

In most cases humans decide on what data to gather. The data can be created by human behaviour, human-system interaction or by system behaviour. Data mostly is a descriptor of the source it comes from and it provides insight into its origin. Data helps to understand systems and therefore it is key in research.

\item [Skills] in general, and digital skills in particular, are fundamental to the health of the UK's financial services sector and its ability to compete and innovate; yet the industry suffers from persistent shortages in Data, Software, Risk, Tech and Cyber roles~\cite{FSSC2023}. 

A diversity of skills in financial products, algorithms, markets, stakeholders and technologies greatly adds to the quality of the data that can be sampled and analysed. 

\item [Codesign]-- that is, the involvement and embedding of users into the design process of a system~-- can lead to the generation of concepts that are more novel and which feature significantly higher levels of user benefit than those generate by in-house professionals, even if more technical imagination may be required to overcome feasibility challenges~\cite{Trischler2017}.  Furthermore, digital technologies are critiqued for the lack of diversity in the involvement of the underbanked, because data is rarely collected from this societal group; a consequent problem is that the resultant products and services can lead to increasing rather than decreasing exclusion to financial services and products. This is why data sharing via OB/OF acts as an enabler to break such barriers; however, more research and progress is required to reach those citizens in society who could potentially benefit the most from being included in the codesign stage of digital innovation processes. 
\item [Collaboration] is a natural  way for academic and non-academic organisations to mutually benefit, and is an essential ingredient of successful codesign projects. In non-collaborative environments diminished innovation outcomes are likely~\cite{Trischler2017}. 

Meaningful collaboration~-- where each party mutually benefits from the delivery of a joint project and sees enhanced outcomes over and above base expectations~-- occurs when each party is genuinely committed to following through on pledges and making use of the deliverables.  Funding is a crucial catalyst to bring parties together initially to work jointly towards co-designing meaningful collaborative research projects.
\item [Ethics] is our last, but certainly not least, enabler. Ethical financial services and technologies are those that contribute to society and prioritise people's wellbeing, that acknowledge that all people are stakeholders in financial services as a basic utility, and that new technology enables products and services should be designed and implemented to avoid harm. Financial services that are designed with ethics at the heart are transparent and trustworthy, are fair and do not discriminate, respects people's privacy and rights as citizens, and foster people's growth. Beyond the ethical considerations of the impacts new technologies and services have on people (consumers, users, stakeholders), taking ethics seriously also means considering the planetary impacts of new systems~-- such as acknowledging and mitigating for the carbon footprints of adopting new data intensive systems. Ethics itself is also enabled by taking into consideration all of the enablers introduced above~-- for example, co-designing with end-users leads to better understandings of people's real needs and how these may be addressed without causing unintended harms; adopting a critical perspective on data means considering the ethical issues around bias, privacy, and the environmental costs of training new models.
\end{description}

\section{Outlook}

\begin{figure}[htbp]
    \includegraphics[width=0.9\textwidth]{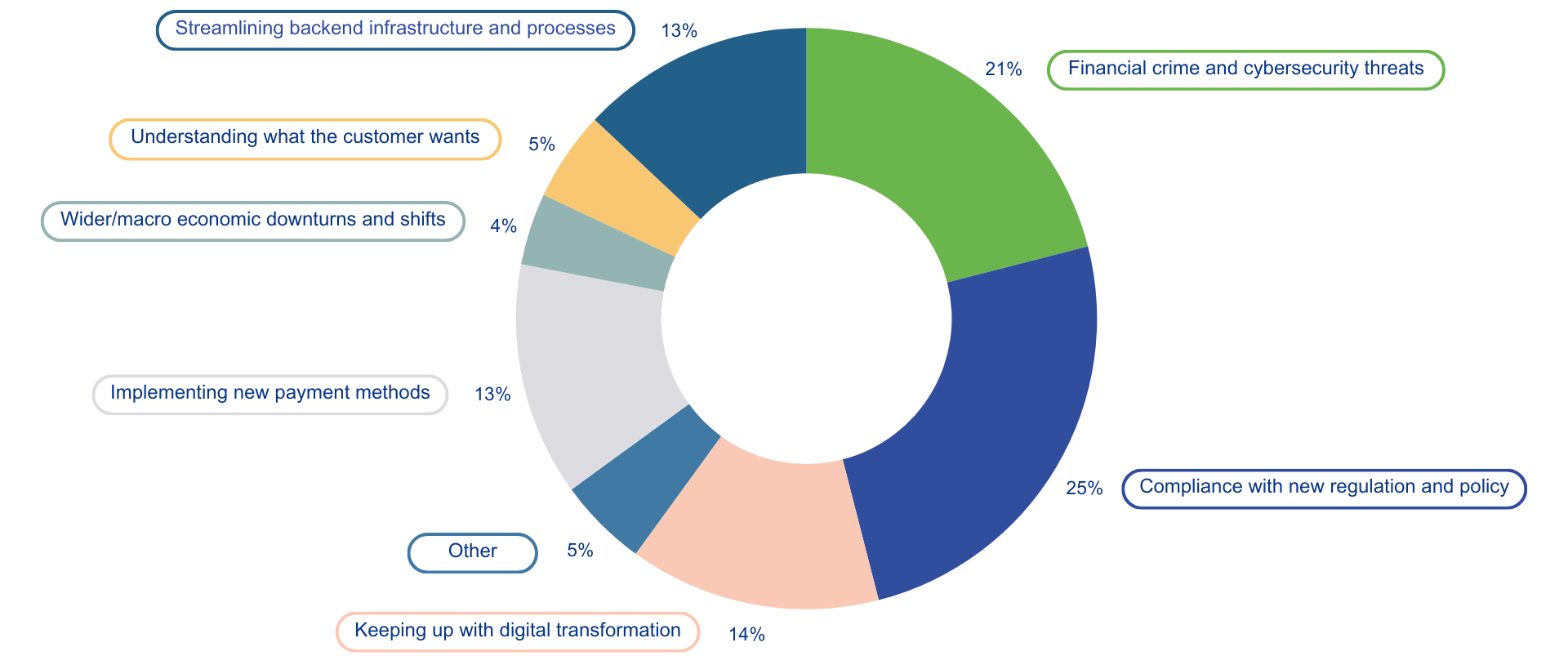}
    \caption{Biggest challenges for payment sector going forward. Source:~\cite{Pay360}}
    \label{fig:paymentchallenges}
\end{figure}

\begin{figure}[htbp]
    \hspace*{8mm}\includegraphics[width=0.86\textwidth]{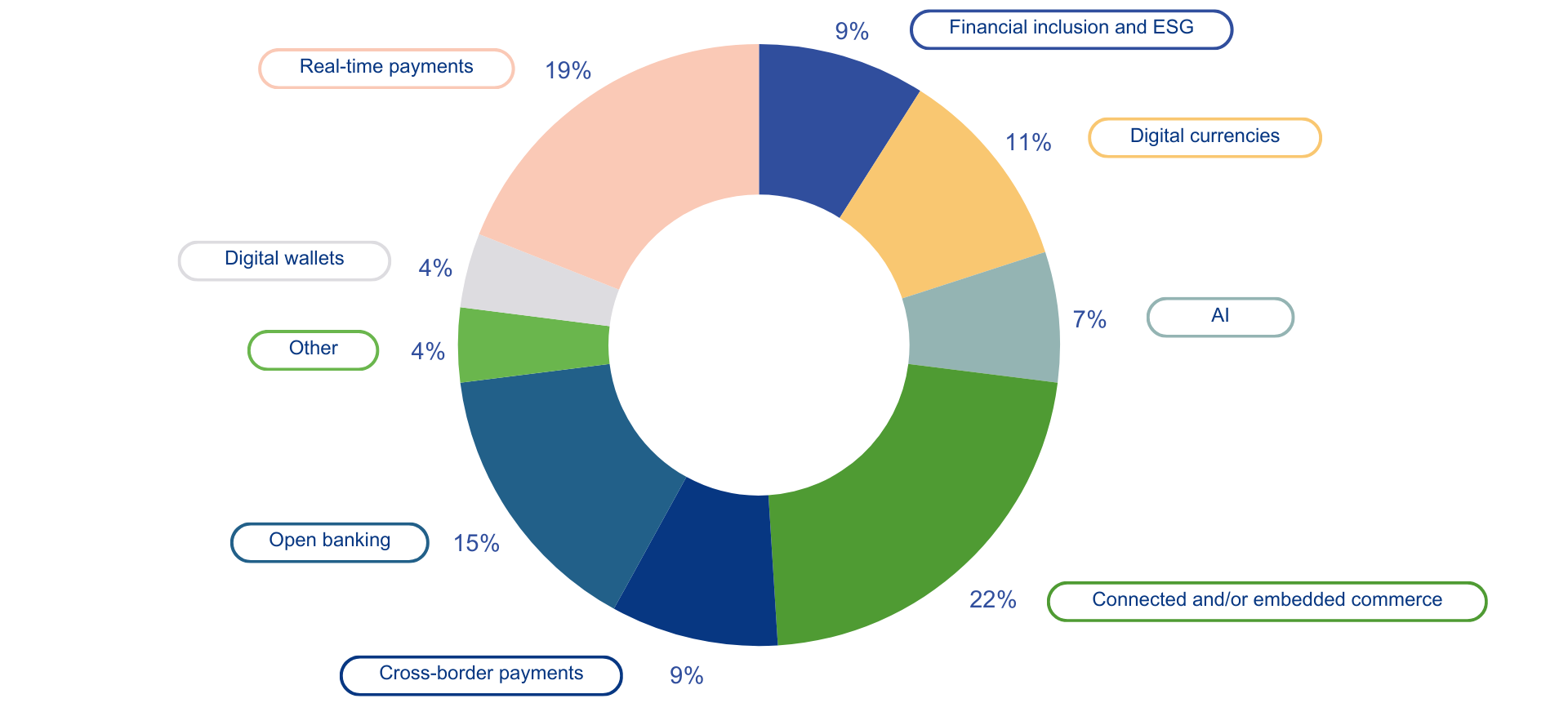}
    \caption{Biggest opportunities for payment sector. Source:~\cite{Pay360}}
    \label{fig:paymentopportunities}
\end{figure}

Looking ahead, we note a good degree of alignment of the UKFin+ research agenda with the challenges and opportunities for the payment sector identified in a recent Payments Association Survey of financial industry professionals, as shown in Figures~\ref{fig:paymentchallenges} and \ref{fig:paymentopportunities}. 

The survey highlighted seven major challenges for the payments sector~-- with compliance, financial crime, keeping pace with the adoption of digital technologies~-- all noted as major areas to be addressed. These map onto several of our real-life financial services challenges: regulatory compliance, coping with game-changing technologies, and financial resilience and stability.

Moving forward, the survey notes eight main opportunities for payments to capitalise on, amongst which connected commerce, real-time payments, digital currencies and open banking all stand out. In the near term, AI and digital assets are also expected to be amongst the priority areas for the innovation policies of the UK and EU\footnote{See \url{https://www.deloitte.com/uk/en/Industries/financial-services/perspectives/regulatory-outlook-sustainable-finance.html}}.

Besides looking to deploy infrastructure for open real-time payments, we expect payment Service Providers will be increasingly looking to cater for alternative payment methods such as Buy-Now-Pay-Later (BNPL) and other credit schemes, and those which use biometrics, digital wallets and distributed ledger technology, which are growing in popularity. Cryptocurrency accounted for only 0.2\% of global ecommerce payment volumes in 2022\footnote{See \url{https://www.fisglobal.com/en/-/media/fisglobal/files/campaigns/global-payments-report/FIS_TheGlobalPaymentsReport2023_May_2023.pdf}} but this is expected to grow substantially, especially as stablecoins gain increasing traction.

While the Payments Association Survey is just one lens into the complex ecosystems of financial services, it highlights multiple starting points for new research projects that require collaboration between academia and industry. The challenges can be linked to opportunities through the research and application areas we highlight in the UKFin+ research agenda. For example, compliance with new regulation is not only further complicated by novel AI tools, but could also be potentially addressed by them: however, new research is needed both at the fundamental technical layer to create LLM tools that provide usable and trusted outputs that can streamline compliance processes, and at the organisational and individual user layer to ensure these tools are understood by compliance experts and are implemented in ways that augment well established workflows and protocols. Another example is growing the diversity of research into understanding what the customer wants by leveraging research in areas such as behavioural finance, understanding user privacy and trust factors, and developing new techniques for doing meaningful co-design at scale and with efficiency. Through addressing these challenges through research in areas like these, we could take better advantage of the  opportunities around open banking, financial inclusion, and digital currencies.

Across all research areas, a shortage of skilled professionals, and the lack of availability of high quality of data is expected to remain problematic. Research proposals addressing these issues would be welcome.

UKFin+ aims to increase collaboration between academic researchers and the economically and socially important financial services industry.  Research funders seeking to consolidate on UKFin+ research funding investment may want to consider focused funding calls where current challenges (pushes) and future opportunities (pulls) identified in the research agenda overlap.  This would enable the seeds sown by UKFin+ funding to grow further, and ensure fintech research funding is directed to where it is most needed in the financial services industry.

We hope the presented research agenda inspires applications not only for UKFin+ funding but also other funding sources including industry-funded projects, thereby supporting innovation and boosting the efficiency, resilience and competitiveness of the UK's financial services sector.





\section*{Acknowledgements}

This work was supported by the Engineering and Physical Sciences Research Council (EPSRC) (Grant number EP/W034042/1). We gratefully acknowledge their support.

We are also grateful to all the attendees of UKFin+ events, especially panelists, presenters and workshop participants, who have helped to shape the UKFin+ research agenda.

Finally, we thank members of the UKFin+ Advisory Board who provided helpful comments and feedback on earlier drafts of this document.

\addcontentsline{toc}{section}{References}
\bibliographystyle{plain}
\bibliography{references}


\end{document}